\documentclass[aps,12pt]{revtex4}
\usepackage{dcolumn}
\usepackage{bm}
\renewcommand{\baselinestretch}{1.5}
\renewcommand{\vec}{\bm}
\begin{document}

\title{Accuracy and efficiency of modern methods
       for electronic structure calculation on
       heavy- and superheavy-element compounds}

\author{Anatoly V.\ \surname{Titov}}
\email{Titov@pnpi.spb.ru}
\homepage{http://qchem.pnpi.spb.ru}
\author{Nikolai S.\ \surname{Mosyagin}}
\author{Timur A.\ \surname{Isaev}}
\author{Aleksander N.\ \surname{Petrov}}
\affiliation{Petersburg Nuclear Physics Institute RAS,
 Gatchina, St.-Petersburg  188300, RUSSIA}

\begin{abstract}
%
  The methods which are actively used for electronic structure calculations of
  low-lying states of heavy- and superheavy-element compounds are briefly
  described.  The advantages and disadvantages of 
  calculations with
  the Dirac-Coulomb-Breit
  Hamiltonian, Huzinaga-type 
  pseudopotential, 
  shape-consistent Relativistic
  Effective Core Potential (RECP) and Generalized RECP are discussed.
  The nonvariational technique of the electronic structure restoration in
  atomic cores after the RECP calculation of a molecule is presented.
  The features of 
   some
  approaches accounting for electron correlation, the
  configuration interaction and coupled cluster methods, are also described.
  The results of calculations on
   E113, E114, U and other heavy-atom systems are presented.
%
\end{abstract}

\maketitle

\section{Introduction}

 High-precision calculations of molecules with heavy and superheavy atoms that
 provide ``chemical accuracy'' (1~kcal/mol or 350~cm$^{-1}$)
 for excitation and dissociation energies of low-lying states are extremely
 time-consuming.  Employing the latest theoretical and
 program
 developments is necessary on the following stages:
\begin{enumerate}
 \item[(A)] selection of an effective spin-dependent Hamiltonian;
 \item[(B)] basis set optimization;
 \item[(C)] appropriate way of accounting for correlation.
\end{enumerate}
 In order to minimize the computational efforts necessary to provide a given
 accuracy in calculation of properties, it is important to achieve the
 equivalent (balanced) level of accuracy in each of these stages in the most
 economical way.
 Moreover, too high accuracy which can be formally attained at the first two
 stages by, e.g.,
%
%
 {\it (a)} employing an effective Hamiltonian, in which inactive core
 electrons are treated explicitly or/and {\it (b)} using a too large basis set
 etc.\ can result in abnormal requirements to computers at the last stage.

 In the present paper, the main attention is paid on items (A) and (C).  The
 Dirac-Coulomb-Breit (DCB) Hamiltonian and the Relativistic Effective Core
 Potential (RECP) method which are widely employed [at stage (A)] are described
 in sections~\ref{sDCB} and \ref{sRECPs}. The Configuration Interaction (CI)
 and Coupled Cluster (CC) methods which are
 most popular in correlation calculations,
 [at stage (C)] are presented in sections~\ref{sCI} and \ref{sCC}.
 In opposite to the density functional approaches, the CI and CC methods
 allow one to study excited electronic states of a given symmetry with high
 level of accuracy.

\section{Dirac-Coulomb(-Breit) Hamiltonian}
\label{sDCB} 

 It is well known that the Dirac-Coulomb (DC) Hamiltonian with the Breit
 interaction and other Quantum ElectroDynamic (QED) corrections taken into
 account 
 can theoretically
 provide a very high accuracy of calculations of heavy atoms and
 heavy-atom molecules.  The DC Hamiltonian has the form (in atomic units $e =
 m = \hbar = 1$, where $e$ and $m$ are the electron charge and mass, $\hbar$ is
 Planck constant):
%
\begin{equation}
     {\bf H}^{\rm DC}\ =\ \sum_{p} {\bf h}^{\rm D}(p)
                    + \sum_{p>q} \frac{1}{r_{pq}} \ ,
 \label{DC}
\end{equation}
 where
 indices $p,q$ run over all the electrons in an atom or molecule,
 $r_{pq}$ is the distance between electrons $p$ and $q$, and
 the one-electron Dirac operator ${\bf h}^{\rm D}$ is
\begin{equation}
   {\bf h}^{\rm D}\ =\ c (\vec{\alpha} \cdot \vec{p})
                 + mc^2 (\beta - 1) + V^{\rm nuc}\ ,
 \label{hD}
\end{equation}
 $c$ is the speed of light, $V^{\rm nuc}$ is the nuclear potential 
 which can include
 the effect of finite nuclear size etc.,\quad
 $\vec{p}{=}{-}i\vec{\nabla}$
 is the electron
 momentum
 operator,
 ${\vec \alpha}, \beta$\ are the $4{\times}4$ Dirac matrices.

 The lowest-order QED correction includes the interelectronic exchange by one
 transverse photon in the Coulomb gauge and leads to so-called
 Dirac-Coulomb-Breit
 Hamiltonian,
\begin{equation}
    {\bf H}^{\rm DCB}\ =\ {\bf H}^{\rm DC} + \sum_{p>q} B_{pq}\ ,
 \label{DCB}
\end{equation}
 where
\begin{equation}
   B_{pq}(\omega_{pq}) =
         -({\vec \alpha}_p{\cdot}{\vec \alpha}_{q})
          \frac{\cos (\omega_{pq} r_{pq})}{r_{pq}}
        + ({\vec \alpha}_p{\cdot}{\vec \nabla}_p)
          ({\vec \alpha}_{q}{\cdot}{\vec \nabla}_{q})
        \frac{\cos (\omega_{pq} r_{pq}){-}1}{\omega_{pq}^2 r_{pq}}
   \ ,
 \label{Bijw}
\end{equation}
 $\omega_{pq}$ designates the frequency of the photon exchanged between
 electrons $p$ and $q$.  A low-frequency expansion of the cosines yields the
 incomplete Breit interaction $B_{pq}(0)$:
\begin{equation}
   B_{pq}(0)\ =\ -{\vec \alpha}_p \cdot {\vec \alpha}_{q}/r_{pq}\ +\
                 \frac{1}{2}\left[{\vec \alpha}_p \cdot {\vec \alpha}_{q} -
                ({\vec \alpha}_p \cdot {\vec r}_{pq})
                ({\vec \alpha}_{q} \cdot {\vec r}_{pq})/r_{pq}^2\right]/r_{pq}\ .
 \label{Bij0}
 \end{equation}
 These terms describe the instantaneous magnetostatic interaction
 and classical retardation of the electric interaction between electrons.
 The contribution from the first term (called Gaunt interaction)
 to transition energies and hyperfine structure (HFS) constants
 can be observed in atomic Dirac-Hartree-Fock (DHF) 
 calculations~\cite{Tupitsyn}
 (Tables~\ref{Pb} and \ref{Tl1}).

 The one-electron basis functions in calculations with the DC(B) Hamiltonian
 are the four-component Dirac spinors.  The DC(B)-based calculations have the
 following disadvantages:
\begin{itemize}
\item
         too many electrons are treated explicitly in heavy-atom systems and
         too large basis set of 
	 Gaussian functions 
	 is required for accurate description
	 of the large number of 
         radial
	 oscillations which valence spinors
	 have in the case of a heavy atom;
\item
         the necessity to work with the four-component Dirac spinors
	 leads to serious complication of calculations
         as compared to the nonrelativistic case.
\end{itemize}

\section{Relativistic Effective Core Potentials}
\label{sRECPs} 

 In calculations on heavy-atom molecules, the DC and DCB Hamiltonians are
 usually replaced by an effective Hamiltonian

\begin{equation}
   {\bf H}^{\rm Ef}\ =\ \sum_{p_v} [{\bf h}^{\rm Schr}(p_v) +
          {\bf U}^{\rm Ef}(p_v)] + \sum_{p_v > q_v} \frac{1}{r_{p_v q_v}} \ ,
 \label{Hef}
\end{equation}
 written only for valence or ``valence-extended''
 (when some outermost core shells are treated explicitly) subspace
 of electrons denoted by indices $p_v$ and $q_v$;
 ${\bf U}^{\rm Ef}$ is an RECP
 operator simulating, in particular, interactions of the explicitly
 treated electrons with those which are excluded from the
 RECP calculation.  In Eq.~(\ref{Hef}),
\begin{equation}
     {\bf h}^{\rm Schr}\ = - \frac{1}{2} {\vec \nabla}^2 + V^{\rm nuc}\
 \label{Schr}
\end{equation}
 is the
 one-electron operator of the nonrelativistic Schr\"odinger Hamiltonian.
 Contrary to the four-component wave function used
 in DC(B) calculations, the pseudo-wave function in the RECP case can be both
 two- and one-component.

\subsection{Huzinaga-type potential}
\label{sHuz-PPs} 

 When forming chemical bonds in heavy-atom molecules, states of core electrons
 are practically unchanged.  To reduce computational efforts in expensive
 molecular calculations, the ``frozen core'' approximation is often employed.

 In order to ``freeze'' core ($c$) spinors, the energy level shift technique
 can be applied. Following Huzinaga, {\it et al.}~\cite{Huzinaga}, one should
 add 
 the effective core operator ${\bf U}^{\rm Ef}_{\rm Huz}$ containing 
 the Hartree-Fock~(HF) field operators, the Coulomb (${\bf J}$) and 
 spin-dependent exchange (${\bf K}$) terms, over these core spinors 
 together with the level shift terms to the one-electron part of the 
 Hamiltonian:
%
\begin{equation}
  {\bf U}^{\rm Ef}_{\rm Huz} =
    ({\bf J{-}K})[\varphi_{n_clj}]\ + \sum_{n_c,l,j}  B_{n_clj}\
          |\varphi_{n_clj} \rangle \langle \varphi_{n_clj}|\
          \quad (\mbox{i.e.}\ \ \varepsilon_{n_clj} \to
                 \varepsilon_{n_clj}{+}B_{n_clj})\ ,
 \label{OC_Fr-1}
\end{equation}
 where $n_c$, $l$ and $j$ are the principal, orbital momentum and total
 momentum quantum numbers, the $B_{n_clj}$ parameters are at least of order
 $|2\varepsilon_{n_clj}|$ and $\varepsilon_{n_clj}$ is the one-electron energy
 of the core spinor $\varphi_{n_{c}lj}$ that is frozen.  Such nonlocal terms
 are needed in order to prevent collapse
 of the valence electrons to the frozen core states.  As it will be shown
 below, all the terms with the frozen core spinors (the level shift operator
 and exchange interactions) can be transformed to the spin-orbit
 representation in addition to the spin-independent Coulomb term.

\subsection{Shape-consistent radially-local RECPs}
\label{ShC-RECPs}

In other RECP versions, the valence spinors are smoothed in the core regions.
Consider the shape-consistent radially-local (or semi-local)
RECP developed by K.~Pitzer's group~\cite{Lee,Christ}.  The nodeless numerical
pseudospinors $\widetilde{\varphi}_{n_vlj}(r)$ are constructed of the large
components 
$f_{n_vlj}(r)$ 
of the
valence ($v$) DHF spinors (one pseudospinor for each $l$ and $j$):
%
\begin{equation}
 \widetilde{\varphi}_{n_vlj}(r) =
 \left\{
  \begin{array}{ll}
   f_{n_vlj}(r)\ ,                            &  r\geq R_{c}\ , \\
   r^{\gamma}\sum_{i=0}^{5}a_{i}r^{i}, &  r<R_{c}\ ,
  \end{array}
 \right.
\end{equation}
%
 where $r$ is the distance between the nucleus and electron.  The matching (or
 core) radius, $R_c$, is chosen near the outermost extremum for the large
 component and the $a_{i}$ coefficients are taken such that the pseudospinors
 are
 normalized,
  smooth and nodeless.  The
 power $\gamma$
 is typically chosen higher than $l+1$ to ensure an efficient ejection of the
 valence electrons from the core region.

 To derive the RECP components $U_{lj}$, the HF equations are inverted for the
 valence
 pseudospinors
 so that $\widetilde{\varphi}_{n_vlj}$ become solutions of the
 nonrelativistic-type HF equations (but with $j$-dependent potentials)
 for a ``pseudoatom'' with removed core electrons~\cite{Goddard}:
\begin{equation}
 U_{lj}(r) = \widetilde{\varphi}_{n_vlj}^{-1}(r)
                \biggl(\frac{1}{2}\frac{d^{2}}{dr^{2}}
                - \frac{l(l{+}1)}{2r^{2}}
                + \frac{Z^*}{r} -
                \widetilde{\bf J}(r) +
                \widetilde{\bf K}(r)
              +   \varepsilon_{n_vlj}  \biggr) \widetilde{\varphi}_{n_vlj}(r)
               \ ,
 \label{U_nlj}
\end{equation}
 where $Z^*=Z-N_c$, $Z$ is the nuclear charge, $N_c$ is the number of excluded
 core electrons, $\widetilde{\bf J}$ and $\widetilde{\bf K}$
 are the Coulomb and exchange operators on the pseudospinors
 $\widetilde{\varphi}_{n_vlj}$,
 $\varepsilon_{n_vlj}$ are their one-electron energies (the same as for the
 original spinors).

 The radially-local RECP operator ${\bf U}^{\rm Ef}_{\rm rloc}$ can be written
 in the form:
\begin{equation}
  {\bf U}^{\rm Ef}_{\rm rloc} =
                   \frac{N_c}{r} +
                     U_{LJ}(r)
                   + \sum\limits_{l=0}^L \sum\limits_{j=|l-1/2|}^{l+1/2}
                   \bigl[U_{lj}(r)-U_{LJ}(r)\bigr] {\bf P}_{lj}\ ,\ \
  {\bf P}_{lj} =
     \sum\limits_{m_j=-j}^j \bigl| ljm_j \bigl\rangle \bigr\langle ljm_j \bigr|\ ,
 \label{URECP}
\end{equation}
%
 where\ $J=L+1/2$, $L = l_c^{max}+1$\ and\ $l_c^{max}$\ is the highest orbital
 momentum of the core spinors, $m_j$ is the projection of the total momentum.

 Using the identities for the ${\bf P}_{lj}$ projectors~\cite{Hafner}:
\begin{equation}
        {\bf  P}_{l,j=l\pm 1/2}\
         =\ \frac{1}{2l{+}1} \Bigl[ \Bigl(l +
            \frac{1}{2} \pm \frac{1}{2}\Bigr)
            {\bf P}_l \pm
          2 {\bf P}_l\
          \vec{l}{\cdot}\vec{s}\ {\bf P}_l \Bigr]\ ,\ \
  {\bf P}_{l} =
     \sum\limits_{m_l=-l}^l \bigl| lm_l \bigl\rangle \bigr\langle lm_l \bigr|\ .
\label{Oper_Pnljs}
\end{equation}
 the RECP operator can be rewritten in the spin-orbit representation,
where $\vec{l}$ and $\vec{s}$ are operators of the orbital and spin
momenta, $m_l$ is the projection of the orbital momentum.

 Similar to Huzinaga-type potentials, the shape-consistent radially-local
 RECPs allow one to exclude chemically inactive electrons already from the
 RECP/SCF stage of calculations.  Moreover, they have the following
 advantages:
\begin{itemize}

\item [1] The oscillations of the explicitly treated spinors are smoothed
     in the core regions of heavy atoms when generating nodeless pseudospinors.
     Therefore, the number of the one-electron Gaussian basis functions may be
     minimized, thus reducing dramatically both the number of two-electron
     integrals and the computational time.

\item [2] The small components of the four-component spinors are
     eliminated and the nonrelativistic kinetic energy operator is used.  The
     RECP method allows one to use a well-developed nonrelativistic technique
     of calculation 
     whereas
     relativistic effects are taken into account with the
     help of spin-dependent semi-local potentials.
     Breit and other two-electron QED interactions can be efficiently treated
     within the one-electron RECPs.

\item [3] In principle, correlations of the explicitly treated electrons
     with those which are excluded from the RECP calculation can be considered
     within ``correlated'' RECP versions.  Reducing the number of explicitly
     correlated electrons with the help of the correlated RECPs is a very
     promising way to minimize efforts when performing high-precision
     molecular calculations.
\end{itemize}

The disadvantages of the semi-local RECPs are:

\begin{itemize}
\item [1] By now, different versions of the radially-local RECPs provide a
      comparable level of accuracy for the same number of the explicitly
      treated electrons.
      It is clear that the explicit inclusion of the outer core electrons into
      the RECP calculation is the way to increase the accuracy.
      However, the extension of the space of these electrons more than some
      limit does not improve the accuracy as is obtained in all our
      calculations with RECPs.
%
      The RECP errors still range up to 1000--3000 cm$^{-1}$ and more even 
      for energies of the dissociation of the lowest-lying states and
      of transition between them.


\item [2] The reliability of the radially-local RECP versions is not high for
      transitions with the excitations in $d,f$-shells in transition metals,
      lanthanides, actinides, etc.


\item [3] Moreover, the direct calculation of such properties as
      electronic densities near heavy nuclei, HFS, and matrix elements of
      other operators singular on heavy nuclei is impossible as a result of
      smoothing the spinors in the core regions of heavy elements.
\end{itemize}

 To overcome the above disadvantages, the Generalized RECP (GRECP) method (see
subsection~3.3)
 and the One-Center Restoration (OCR) procedures
 (see 
subsection~3.4)
were developed.

\subsection{Generalized RECP}
\label{sGRECP}

 It was shown in paper~\cite{Mitr} that a requirement for pseudospinors to be
 nodeless is not necessary to generate the shape-consistent RECP components.
 In the case of pseudospinors with nodes, the RECP components are singular
 because division by zero appears in Eq.~(\ref{U_nlj}). This problem is
 overcome in the GRECP method by interpolating the potentials in the vicinity
 of these nodes.  It was shown both theoretically and computationally that the
 interpolation errors are small enough.  This allows one to
 generate different potentials, $U_{n_clj}$ and $U_{n_vlj}$, for outer core
 and valence pseudospinors, unlike the conventional RECP approach.

 The GRECP operator is written in the form~\cite{Tup}:
\begin{eqnarray}
 {\bf U}^{\rm GRECP} & = & \frac{N_c}{r} + U_{n_vLJ}(r)
                   + \sum_{l=0}^L \sum_{j=|l-1/2|}^{l+1/2}
                   \bigl[U_{n_vlj}(r)-U_{n_vLJ}(r)\bigr]
                   {\bf P}_{lj}                             \nonumber\\
             & + & \sum_{n_c} \sum_{l=0}^L \sum_{j=|l-1/2|}^{l+1/2}
                   \Bigl\{\bigl[U_{n_clj}(r)-U_{n_vlj}(r)\bigr]
                   \widetilde{\bf P}_{n_clj} + \widetilde{\bf P}_{n_clj}
                   \bigl[U_{n_clj}(r)-U_{n_vlj}(r)\bigr]\Bigr\}     \nonumber\\
             & - & \sum_{n_c,n_c'} \sum_{l=0}^L \sum_{j=|l-1/2|}^{l+1/2}
                   \widetilde{\bf P}_{n_clj}
                   \biggl[\frac{U_{n_clj}(r)+U_{n_c'lj}(r)}{2}-U_{n_vlj}(r)\biggr]
                   \widetilde{\bf P}_{n_c'lj},
 \label{UGRECP}
\end{eqnarray}
\[
  \widetilde{\bf P}_{n_clj} =
     \sum\limits_{m_j=-j}^j \bigl| \widetilde{n_cljm_j} \bigl\rangle
     \bigr\langle \widetilde{n_cljm_j} \bigr|\ .
\]
 The new non-local terms (the second and third lines in the above equation)
 were added to the conventional semi-local RECP operator. These terms take
 into account the difference between the effective potentials acting on the
 outer core and valence electrons with the same $l$ and $j$ quantum numbers.

 The GRECP method allows one to improve accuracy of calculations by regular
 manner when including more outer core shells explicitly into the GRECP
 calculations.  More details on the GRECP method can be found
 in~\cite{Theor,Obzor}.  To compare different effective potential versions by
 accuracy, we carried out both all-electron calculations with the DC
 Hamiltonian and calculations with RECPs of different groups.  The RECP errors
 in reproducing the DHF all-electron results are studied in \cite{Theor,Obzor}
 etc.  One can see from our atomic HF calculations~\cite{Latajka} and
 correlation calculations on the Hg~\cite{Hg} and Pb~\cite{Pb} atoms, that the
 accuracy of the GRECP is
 up to an order of magnitude higher than that of the other tested RECPs even
 for the
 cases when the same number of 
 outermost core shells is treated
 explicitly.

 Results for the eka-thallium atom (E113) are presented in Table~\ref{E113}.
 The GRECP errors are collected into two groups. The errors for transitions
 without change in the occupation number of the $6d$ shell are rather small.
 The errors for transitions with change in the occupation number of the $6d$
 shell are about 400~cm$^{-1}$. The latter errors have a systematic nature and
 are connected with the fact that the $6d$ shell in the present GRECP version
 is described with the help of nodeless pseudospinors. Of course, these errors
 can be reduced significantly if one includes the $5d$ electrons explicitly in
 the GRECP calculations.  The Self-Consistent (SfC) RECP method was suggested
 in \cite{SfC,Theor}, it allows one to minimize the above mentioned errors
 without extension of space of explicitly treated electrons.  New terms with
 an operator of the occupation number of the outermost $d$ (or $f$) shell are
 added to the RECP operator. This method is most optimal for studying
 compounds of transition metals, lanthanides, and actinides.  The comparison
 of accuracy of different RECP versions in calculations on the uranium atom
 can be found in Table~\ref{U_SfC1} and in papers~\cite{Theor,SfC}.
 
 A technique for the ``Correlated'' GRECP (CGRECP) generation was proposed
 in~\cite{Theor} and essential improvements in this technique were made
 in~\cite{Obzor}.  The CGRECP for mercury was generated in the framework of
 relativistic CC~(RCC) method~\cite{Kaldor}. The GRECP components were
 constructed for $5s,5p,5d,6s,6p,6d,5f,5g$ electrons of the Hg atom.  The
 $5s,5p$ pseudospinors are ``frozen'' in calculations with this CGRECP (in
 fact, they are completely excluded from  the calculations with the help of the
 level shift technique (see 
 section~3.1
 and Refs.~\cite{Theor,TlH}))
 and only 12 external electrons of the Hg atom should be explicitly correlated
 instead of 34 ones in the case of the DC calculations because the correlations
 for the $4f,5s,5p$ electrons and between the $4f,5s,5p$ and $5d,6s,6p$
 electrons are taken into account by the ``correlated'' GRECP.  It allows one
 to reduce drastically the computational efforts necessary for the ``chemical''
 accuracy in calculations of mercury compounds.  Results of our test
 calculations are presented in Table~\ref{CGRECP}.
 One can see that energies of the one-electron excitations 
 in calculations of the Hg atom with the all-electron DC Hamitonian for 34
 external electrons correlated explicitly by the RCC method (DC/34e-RCC) can be
 reproduced with the accuracy withing 270~cm$^{-1}$ in the 12e-RCC calculations when
 using the present CGRECP.
 Contribution of the core correlation effects can be seen from comparison of
 the DC/12e-RCC and DC/34e-RCC results or of the GRECP and CGRECP results.

\subsection{Nonvariational One-Center Restoration
 of electronic structure in cores of heavy-atoms in a molecule (NOCR)}
\label{sNOCR} 

 In the valence
 region, the electronic density obtained from the two-component GRECP
 (pseudo)wave function very accurately reproduces the corresponding
 all-electron four-component density.  In the
 core region, the pseudospinors are smoothed, so that
 the electronic density with the (pseudo)wave function is not correct.


  The following restoration scheme was developed (see \cite{TlF,Var} and
  references):
\begin{itemize}
\item  Generation of equivalent basis sets of atomic (one-center)
       four-component spinors
 $\left\{ \left( \begin{array}{c} f_{nlj}(r)\chi_{ljm_j} \\
  g_{nlj}(r)\chi_{l'jm_j} \\ \end{array} \right) \right\}$
(where $f_{nlj}$, $g_{nlj}$ are the radial parts, $\chi_{ljm_j}$ are
the spin-angular parts of the atomic Dirac spinors and $l'{=}2j{-}l$)
 and two-component pseudospinors $\{\tilde f_{nlj}(r)\chi _{ljm_j}\}$
 by atomic
 finite-difference
(numerical)
 all-electron DHF and two-component GRECP/HF calculations of the same valence
 configurations of the
 atom and its ions.
%
%

\item
 The molecular pseudospinorbitals $\tilde {\phi} _{i}$ are then expanded in
 the basis set of the one-center two-component atomic pseudospinors (for
 $r{\le}R_c^{\rm rest}$, where $R_c^{\rm rest}{\ge}R_c$),
\begin{equation}
    \tilde {\phi} _{i}({\bf x}) \approx
    \sum_{l=0}^{L_{max}}\sum_{j=|l-1/2|}^{l+1/2} \sum_{n,m_j}
    c_{nljm_j}^{i}\tilde f_{nlj}(r)\chi _{ljm_j}\ ,
 \label{expansion}
\end{equation}
 where ${\bf x}$ denotes spatial and spin variables.
%
\item
 Finally, the atomic two-component pseudospinors are replaced by the
 equivalent four-component spinors in the molecular basis and the expansion
 coefficients $c_{nljm_j}^{i}$ from Eq.~(\ref{expansion}) are preserved:
%
\begin{equation}
{\phi} _{i}({\bf x}) \approx
    \sum_{l=0}^{L_{\rm max}}\sum_{j=|l-1/2|}^{l+1/2} \sum_{n,m_j}
    c_{nljm_j}^{i}
    \left(
    \begin{array}{c}
    f_{nlj}(r)\chi _{ljm_j}\\
    g_{nlj}(r)\chi _{l'jm_j}
    \end{array}
    \right)\ .
  \label{restoration}
\end{equation}
\end{itemize}

 The molecular four-component spinors constructed this way are orthogonal to
 the inner core spinors of the heavy atom, as the atomic basis functions used
 in Eq.~(\ref{restoration}) are generated with the inner core electrons
 treated as frozen.
 The properties described by the operators singular close to (heavy) nuclei
 are calculated with the restored bispinors ${\phi} _{i}$ .  More advanced
 technique of the variational restoration is proposed in \cite{Var}.

\section{Configuration Interaction}
\label{sCI} 

%
%
%
%
%

%

 The many-electron wavefunction $\Psi^{CI}$ in the CI method is presented by a
 linear combination of determinants $D_I$
\begin{equation}
   \Psi^{\rm CI}=\sum_I C^{\rm CI}_I D_I\ ,
\end{equation}
 $C^{\rm CI}_I$ are some numbers (CI coefficients).  In turn, each 
 determinant is an anti-symmetric
 production of $N$ one-electron basis functions where $N$ is the number of
 electrons in the considered system.  The CI equations are written as
\begin{equation}
   \sum_J H_{IJ} C^{\rm CI}_J =E^{\rm CI} C^{\rm CI}_I\ ,
\end{equation}
 where $H_{IJ}$ are Hamiltonian matrix elements in the basis set of the
 determinants and $E^{\rm CI}$ is the CI energy.  To find the coefficients and the
 energy in the CI method, one should diagonalize the Hamiltonian matrix.

 If all 
 the
 possible determinants are considered then the method (called Full-CI)
 will provide the ``exact'' solution in the framework of a given one-electron
 basis set and an employed Hamiltonian.  However, requirements to the
 computational resources in the Full-CI case are usually so huge that such
 calculations are practically impossible for systems of interest except the
 cases of very small numbers of correlated electrons and basis functions.  In
 almost all the CI calculations, only some selected (the most important)
 determinants are explicitly considered.  To take into account the effect of
 the unselected determinants, various semi-empirical corrections (e.g., the
 Davidson correction \cite{Davidson}) can be employed.  In precise
 calculations, the number of selected determinants reaches a few millions and
 more, therefore a very large Hamiltonian matrix should be diagonalised.  The
 iterative diagonalization (Davidson) method is then used to obtain a few
 low-lying roots of this matrix.

%

 There are two main categories of the CI method~\cite{MRDCI}:
\begin{itemize}
\item
     ``Conventional CI'':
      the Hamiltonian matrix elements are calculated once and saved in memory,
\item ``Direct CI'': only those Hamiltonian matrix elements are calculated at
      each step of the diagonalization procedure which are required at the
      moment.
\end{itemize}

The CI method has the following advantages:
\begin{itemize}
\item  [1] simplicity of the method, solutions are always exist
       independently of the number of open shells;
\item  [2] it well describes ``static'' (avoided crossing of terms)
       and ``nondynamic'' electron correlations.
\end{itemize}

The disadvantages of the CI method are:
\begin{itemize}
\item  [1] it is badly working for large number of correlated electrons
       (when semi-empirical corrections on unselected determinants are large);
\item  [2] unsmoothness of potential curves is a result of selection of
       determinants by some thresholds;
\item  [3] the above semi-empirical energy corrections cannot be used
       when calculating other than spectroscopic properties.
\end{itemize}


\section{The Coupled-Cluster Approaches}
\label{sCC} 

%
 The complete space of $\{D_I\}$ is divided into two subspaces:
\begin{itemize}
\item[${\cal M}_0$]
    , model space, consists of small number~($M$) of the most important
    determinants $\{ D_m \}_{m{=}1}^M$ to
    describe static and nondynamic correlations,
    which are taken into account
    exactly on ${\cal M}_0$;

\item[${\cal M}_0^{\perp}$]
    , rest of space (usually very large), is included approximately
    to account for dynamic correlations
 (i.e.\ correlations at small interelectronic distances, ``Coulomb holes'').
\end{itemize}


 The eigenstates of interest are presented as
\begin{equation}
   |\Psi^{\rm CC}\rangle\ = \sum\limits_{m{=}1}^M C_m {\bf exp}[T^{(m)}]
   |D_m\rangle\ ,
 \label{Psi_k}
\end{equation}
 where $T^{(m)} \equiv T_1^{(m)}{+}T_2^{(m)}{+}\dots$\quad is the
 cluster operator:
\begin{equation}
 \left\{
  \begin{array}{l}
   T_1^{(m)} = \sum\limits_{i,a} \left\{ {{\bf a}_a}^+ {\bf a}_i \right\}
                t_{i.a}^{(m)}\ , \vspace{0mm}\\
   T_2^{(m)} = \frac{1}{2}\sum\limits_{ij,ab}\left\{ {{\bf a}_b}^+ {{\bf a}_a}^+
               {\bf a}_j {\bf a}_i \right\} t_{ij,ab}^{(m)}\ , \vspace{-5mm}\\
   \dots\ . \vspace{0mm}\\
  \end{array}
 \right.
\end{equation}
%
 where ${\bf a}_a^+$ and ${\bf a}_i$ are the creation and annihilation
 operators (their combination ${\bf a}_a^+ {\bf a}_i$ will replace the $i$-th
 one-electron state in the determinant by the $a$-th one).  The
 coefficients $\{ t_{i,a}^{(m)}, t_{ij,ab}^{(m)} \}$, etc.\ are called the
 cluster amplitudes and
 are calculated solving
 Bloch equations:
\begin{equation}
 {\bf UHU} = {\bf HU}\ ,\qquad
 ({\bf U} \equiv
     \sum\limits_{m{=}1}^M {\bf exp}[T^{(m)}] |D_m\rangle\langle D_m|)\ .
\end{equation}
%
 The coefficients $C_m$ and final energy $E^{\rm CC}$ are obtained
 from diagonalization of some effective Hamiltonian
 ${\bf H}^{\rm eff}$
 on the model space:
\begin{equation}
{\bf H}^{\rm eff} \sum\limits_{m{=}1}^M C_m |D_m\rangle =
      E^{\rm CC}\sum\limits_{m{=}1}^M C_m |D_m\rangle\ ,  \ \
 ({\bf H}_{nm}^{\rm eff} \equiv
 \langle D_n|
 ({\bf exp}[-T^{(m)}]{\bf H}{\bf exp}[T^{(m)}])
 |D_m\rangle)\ .
\end{equation}

 If all the $T_k^{(m)}$ are considered in the $T^{(m)}$ operator then the CC
 method is equivalent to the Full-CI one. However, in practical calculations,
 the third and following terms in $T^{(m)}$ (three-body and higher order
 cluster amplitudes) are usually neglected. Such a CC version is called CC-SD.
 There are three basic CC categories~\cite{Paldus}:
\begin{itemize}
 \item One-state or state-selective; 
 \item Fock-space or valence universal methods;
 \item Hilbert-space or state-universal approaches.
\end{itemize}

 The CC method has the following advantages:
\begin{itemize}
\item [1] It is the size-extensive method, i.e.\ the energy of the system
     is scaled properly with increase in the number of electrons (whereas the
     CI method is not size-extensive in a general case).

\item [2] The CC-SD method takes into account the contributions not only
     from the determinants 
     obtained from the model space by applying 
     the $(1{+}T_1^{(m)}{+}T_2^{(m)})$ operator 
     but also approximately from all the
     rest determinants (whereas the CI method with the same number of unknown
     coefficients does not).

\item [3] The CC method is one of the best methods for accounting the
     dynamic correlation.
\end{itemize}

 The disadvantages of the CC method are:
\begin{itemize}
\item [1] This is a nonvariational method, i.e.\ the CC energy is not an
     upper bound to the exact energy of the system (whereas the CI energy is).

\item [2] The CC equations are nonlinear and the effective Hamiltonian
     is non-Hermitian.

\item [3] Intruder states (i.e.\ such states from the
     ${\cal M}_0^{\perp}$ subspace, which are lying within the ${\cal M}_0$
     subspace energy span) destroy the convergence of the CC iterations.
     Alleviation the problem is in using:

\vspace{-3mm}
\begin{itemize}
\vspace{-3mm}
\item Incomplete model space procedures;

\vspace{-3mm}
\item Energy shifting, RLE~\cite{RLE}, DIIS~\cite{DIIS1,DIIS2},
     IPM~\cite{IPM} procedures.
\end{itemize}

\end{itemize}

\section{Some practical calculations}
\label{Prac} 

 Calculations of the spectroscopic constants for the ground and lowest excited
 states of the HgH molecule and for the ground state of the HgH$^+$ ion were
 carried out with the help of the GRECP and
 RCC~\cite{Kaldor}
 methods in
 \cite{HgH}. The results are within a few mbohr from the experimental data
 for bond lengths, tens of wave numbers for excitation energies and
 vibrational frequencies. It is demonstrated that the triple cluster
 amplitudes for the 13 outermost electrons and corrections for the Basis Set
 Superposition Errors (BSSE) \cite{BSSE,BSSE1} are necessary to obtain
 accurate results for this molecule.
 The accurate GRECP/CI calculations of the spectroscopic constants for the
 ground state of the TlH molecule are presented in~\cite{TlH}, in which the
 reliability of the semi-empirical energy corrections is in particular
 investigated.

 The NOCR scheme was applied in the 
 GRECP/RCC 
 calculations of the $P,T$-odd
 properties for the TlF molecule \cite{TlF}.  The corresponding GRECP/HF/NOCR
 results are in good agreement with the all-electron DHF results of other
 groups. Inclusion of electron correlation has changed the values on 20\%.
 The previous NOCR version was employed in the GRECP calculations of the
 $P,T$-odd parameters and HFS constants for the YbF~\cite{YbF1,YbF} and
 BaF~\cite{BaF} molecules.  A reasonable agreement with the experimental data
 for the HFS constants was attained.  It was demonstrated that the
 spin-correlation effects of the unpaired electron with the deeply-lying outer
 core $5s$ and $5p$ shells 
 of heavy atom
 should be taken into account in order to perform
 accurate calculations of the HFS and $P,T$-odd constants.

 The authors are grateful to the U.S.\ Civilian Research \& Development
 Foundation for the Independent States of the Former Soviet Union (CRDF) for
 the Grant No.\ RP2--2339--GA--02. 
A.T. and N.M. were supported in part by
Scientific Program of St.Petersburg Scientific Center of RAS. T.I. is grateful
to INTAS grant YSF 2001/2-164 for financial support. A.P. is grateful to Ministry 
of education of Russian Federation (grant PD02-1.3-236) and  to St-Petersburg 
Committee of science (grant PD02-1.3-236 ). This work was also supported in
part by RFBR (grant no. 03-03-32335).
%


%

\renewcommand{\baselinestretch}{1}
\clearpage

\begin{table}
\setcaptionmargin{0mm} \onelinecaptionsfalse
\captionstyle{flushleft}
\caption{\label{Pb}
   Transition energies of the Tin ($Z{=}50$), Lead ($Z{=}82$) and
   Eka-lead ($Z{=}114$) atoms calculated by the DHF method with Coulomb
   and Coulomb-Gaunt two-electrons interaction for states with the $ns^2np^2$
   configuration (in cm$^{-1}$).}
\begin{tabular}{@{}lcdddd}
 configuration  &    J    & \text{DC} & \text{DCG} &  \text{absolute} &  \text{relative\ (\%)}  \\
                &         &        &          & \text{difference}&  \text{difference}   \\
\multicolumn{6}{c}{Tin} \\
\hline
 $(5s^2_{1/2}5p^2_{1/2})           $  & 0 &  3113  &  3153  &   40 &  1.3 \\
 $(5s^2_{1/2}5p^1_{1/2}5p^1_{3/2}) $  & 1 &     0  &     0  &    0 &  0   \\
 $(5s^2_{1/2}5p^1_{1/2}5p^1_{3/2}) $  & 2 &  5143  &  5139  &   -4 & -0.1 \\
 $(5s^2_{1/2}5p^2_{3/2})           $  & 2 &  5941  &  5893  &  -48 & -0.8 \\
 $(5s^2_{1/2}5p^2_{3/2})           $  & 0 & 15873  & 15820  &  -53 & -0.3 \\
\multicolumn{6}{c}{Lead} \\
\hline
 $(6s^2_{1/2}6p^2_{1/2})           $  & 0 & 0      &  0      &     0  &    0  \\
 $(6s^2_{1/2}6p^1_{1/2}6p^1_{3/2}) $  & 1 & 4752   &  4644   &  -108  &  -2.3 \\
 $(6s^2_{1/2}6p^1_{1/2}6p^1_{3/2}) $  & 2 & 9625   &   9514  &  -111  &  -1.2 \\
 $(6s^2_{1/2}6p^2_{3/2})           $  & 2 & 18826  &  18592  &  -234  &  -1.2 \\
 $(6s^2_{1/2}6p^2_{3/2})           $  & 0 & 28239  &  27995  &  -244  &  -0.9 \\
\multicolumn{6}{c}{Eka-lead} \\
\hline
 $(7s^2_{1/2}7p^2_{1/2})           $  & 0 &       0  &      0  &     0  &   0   \\
 $(7s^2_{1/2}7p^1_{1/2}7p^1_{3/2}) $  & 1 &  27198   &  26806  &  -392  &  -1.4 \\
 $(7s^2_{1/2}7p^1_{1/2}7p^1_{3/2}) $  & 2 &  30775   &  30391  &  -384  &  -1.2 \\
 $(7s^2_{1/2}7p^2_{3/2})           $  & 2 &  66068   &  65225  &  -843  &  -1.3 \\
 $(7s^2_{1/2}7p^2_{3/2})           $  & 0 &  74527   &  73674  &  -853  &  -1.1 \\
\hline
\end{tabular}
\end{table}

\begin{table}
\setcaptionmargin{0mm} \onelinecaptionsfalse
\captionstyle{flushleft}
\caption{\label{Tl1}
 HFS constants in the Indium ($Z{=}49$), Thallium ($Z{=}81$)
 and Eka-thallium ($Z{=}113$) atoms calculated by the DHF method with
 Coulomb and Coulomb-Gaunt
 interaction for different configurations (in MHz).}
\begin{tabular} {ldddd}
 configuration  & \text{DC} & \text{DCG} &  \text{absolute}   &   \text{relative\ (\%)}\\
                &          &           &  \text{difference} & \text{difference}  \\
\multicolumn{5}{c}{Indium} \\
\hline
 $(5s^2_{1/2}5p^1_{1/2}) $ &  1913      &  1900     &  -13       &  -0.7 \\
 $(5s^2_{1/2}5p^1_{3/2}) $ &   288      &   287     &   -1       &  -0.3 \\
 $(5s^2_{1/2}5d^1_{3/2}) $ &     4.41   &     4.40  &   -0.01    &  -0.2 \\
 $(5s^2_{1/2}5d^1_{5/2}) $ &     1.88   &     1.88  &    0.0     &   0.0 \\
 $(5s^2_{1/2}6s^1_{1/2}) $ &  1013      &  1011     &   -2       &  -0.2 \\
\multicolumn{5}{c}{Thallium} \\
\hline
 $(6s^2_{1/2}6p^1_{1/2}) $ &  18918     &  18691     &  -227      &  -1.2 \\
 $(6s^2_{1/2}6p^1_{3/2}) $ &   1403     &   1391     &   -12      &  -0.9 \\
 $(6s^2_{1/2}6d^1_{3/2}) $ &     20.8   &     20.8   &     0.0    &   0.0 \\
 $(6s^2_{1/2}6d^1_{5/2}) $ &      8.72  &      8.70  &    -0.02   &  -0.2 \\
 $(6s^2_{1/2}7s^1_{1/2}) $ &   7826     &   7807     &   -19      &  -0.2 \\
\multicolumn{5}{c}{Eka-thallium\footnote{
The magnetic moment $\mu_N$ and spin $I$ for the Eka-thallium nucleus were
       taken as those for Thallium.
The presented results can be easily recalculated as only the proper values
of $\mu_N$ and $I$ are known because they just include
the $\mu_N/I$ coefficient.
       }}  \\
\hline
 $(7s^2_{1/2}7p^1_{1/2}) $ &  150168    &  147538    &   -2630   &  -1.8 \\
 $(7s^2_{1/2}7p^1_{3/2}) $ &    2007    &    1983    &     -24   &  -1.2 \\
 $(7s^2_{1/2}7d^1_{3/2}) $ &    34.3    &    34.2    &    -0.1   &  -0.3 \\
 $(7s^2_{1/2}7d^1_{5/2}) $ &    13.5    &    13.5    &     0.0   &   0.0 \\
 $(7s^2_{1/2}8s^1_{1/2}) $ &   28580    &   28473    &    -107   &  -0.4 \\
\hline
\end{tabular}
\end{table}

\begin{table}
\setcaptionmargin{0mm} \onelinecaptionsfalse
\captionstyle{flushleft}
\caption{\label{E113}
Transition energies between low-lying configurations of
the eka-thallium (E113) atom derived from all-electron calculations
and the errors of their reproducing in calculations with different
RECP versions. All values are in cm$^{-1}$.}
\begin{tabular}{|l|r|r|r|r|}
\hline
                                  &            &              &           & 21e-             \\
		                  & All-el.    & Gaunt        & 21e-      & RECP             \\
		                  & DHFG\footnote{All-electron Dirac-Hartree-Fock-Gaunt
	(DHFG) calculation with Fermi nuclear charge distrtibution for $A=297$.}
                                               & contrib.\    & GRECP\footnote{GRECP generated in the present work from DHFG calculation.}
					                      & of Nash          \\
				  &            &              &           & {\it et al.}\footnote{RECP from~\cite{Nash}  (generated from DHF calculation without Gaunt iteraction).}  \\
\cline{2-5}
 Configuration                    & Transition & DHFG \hfill  & \multicolumn{2}{|c|}{ Absolute errors } \\
	                          & energies   & \hfill - DHF & \multicolumn{2}{|c|}{                 } \\
\hline
 $6d_{3/2}^4 6d_{5/2}^6 7s_{1/2}^2 7p_{1/2}^1 (J=1/2) \rightarrow    $  &           &        &                 &             \\
 $6d_{3/2}^4 6d_{5/2}^6 7s_{1/2}^2 7p_{3/2}^1 (J=3/2)                $  &    25098  &  347   &            -23  &         282 \\
 $6d_{3/2}^4 6d_{5/2}^6 7s_{1/2}^2 8s_{1/2}^1 (J=1/2)                $  &    34962  &  374   &              0  &        -186 \\
 $6d_{3/2}^4 6d_{5/2}^6 7s_{1/2}^2 6f^1 (nonrel.av.)                 $  &    50316  &  395   &              6  &         148 \\
 $6d_{3/2}^4 6d_{5/2}^6 7s_{1/2}^2 5g^1 (nonrel.av.)                 $  &    52790  &  395   &              6  &         148 \\
 $6d_{3/2}^4 6d_{5/2}^6 7s_{1/2}^2 7d^1 (nonrel.av.)                 $  &    45215  &  395   &              6  &         161 \\
 $6d_{3/2}^4 6d_{5/2}^6 7s_{1/2}^2 (J=0)                             $  &    57180  &  395   &              6  &         148 \\
 $6d_{3/2}^4 6d_{5/2}^6 7s_{1/2}^1 7p_{1/2}^2 (J=1/2)                $  &    61499  &  -60   &             32  &        4830 \\
 $6d_{3/2}^4 6d_{5/2}^6 7s_{1/2}^1 7p_{1/2}^1 7p_{3/2}^1 (rel.av.)   $  &    83177  &  248   &             -4  &        5177 \\
 $6d_{3/2}^4 6d_{5/2}^6 7s_{1/2}^1          7p_{3/2}^2 (rel.av.)     $  &   112666  &  624   &             -9  &        5729 \\
 $6d_{3/2}^4 6d_{5/2}^6 7s_{1/2}^1 7p_{1/2}^1 (rel.av.)              $  &   115740  &  268   &             -2  &        5161 \\
 $6d_{3/2}^4 6d_{5/2}^6 7s_{1/2}^1 7p_{3/2}^1 (rel.av.)              $  &   149526  &  678   &            -10  &        5811 \\
 $6d_{3/2}^4 6d_{5/2}^6 7s_{1/2}^1 (J=1/2)                           $  &   234385  &  796   &             -4  &        6151 \\
 \hline                                                                                        
 $6d_{3/2}^4 6d_{5/2}^5 7s_{1/2}^2 7p_{1/2}^2 (J=5/2)                $  &    47410  & -778   &            403  &       -2389 \\
 $6d_{3/2}^4 6d_{5/2}^5 7s_{1/2}^2 7p_{1/2}^1 7p_{3/2}^1 (rel.av.)   $  &    74932  & -424   &            341  &       -2089 \\
 $6d_{3/2}^4 6d_{5/2}^5 7s_{1/2}^2            7p_{3/2}^2 (rel.av.)   $  &   110435  &   -6   &            306  &       -1556 \\
 $6d_{3/2}^3 6d_{5/2}^6 7s_{1/2}^2 7p_{1/2}^2 (J=3/2)                $  &    78862  & -416   &            375  &       -2272 \\
 $6d_{3/2}^3 6d_{5/2}^6 7s_{1/2}^2 7p_{1/2}^1 7p_{3/2}^1 (rel.av.)   $  &   104097  &  -86   &            405  &       -1968 \\
 $6d_{3/2}^3 6d_{5/2}^6 7s_{1/2}^2            7p_{3/2}^2 (rel.av.)   $  &   137083  &  306   &            473  &       -1436 \\
 $6d_{3/2}^4 6d_{5/2}^5 7s_{1/2}^2 7p_{1/2}^1 (rel.av.)              $  &   110139  & -407   &            380  &       -2317 \\
 $6d_{3/2}^4 6d_{5/2}^5 7s_{1/2}^2 7p_{3/2}^1 (rel.av.)              $  &   150116  &   45   &            338  &       -1679 \\
 $6d_{3/2}^3 6d_{5/2}^6 7s_{1/2}^2 7p_{1/2}^1 (rel.av.)              $  &   139841  &  -65   &            439  &       -2184 \\
 $6d_{3/2}^3 6d_{5/2}^6 7s_{1/2}^2 7p_{3/2}^1 (rel.av.)              $  &   177157  &  361   &            506  &       -1541 \\
 $6d_{3/2}^4 6d_{5/2}^5 7s_{1/2}^2 (J=5/2)                           $  &   239509  &  158   &            408  &       -1603 \\
 $6d_{3/2}^3 6d_{5/2}^6 7s_{1/2}^2 (J=3/2)                           $  &   267208  &  481   &            579  &       -1431 \\
\hline                                                                               
\end{tabular}
\end{table}

\begin{table}
\setcaptionmargin{0mm} \onelinecaptionsfalse
\captionstyle{flushleft}
\caption{\label{U_SfC1}
Transition energies between states of U
(averaged over nonrelativistic configurations)
derived from all-electron DHF calculations and
the errors of their reproducing in calculations with different RECP versions.
All values are in cm$^{-1}$.}
\begin{tabular}{|l|r|r|r|r|r|r|r|r|}
\hline
                              &          &   RECP of  &  Energy-   &            & Quadratic  &\multicolumn{2}{|c|}{``Frozen} \\
                              &   DHF    &   Ermler   &  adjusted  & SfC        & SfC        &\multicolumn{2}{|c|}{  core''} \\
                              &          &{\it et al.}~\cite{Ermler1}
                                         &   PP\footnote{PseudoPotential (PP) from~\cite{Kuchle1} (generated from all-electron calculation
					 in the framework of Wood-Boring~\cite{Wood} approximation).}
					                           & GRECP      & GRECP      &    ($f^3$) &   ($f^2$)      \\
\hline
Num.\  of el-ns               &  All     &    14      &    32      &    24      &    24      &    24      &    24          \\
\hline
       Conf.\                 &Tr.energy &                           \multicolumn{6}{|c|}{Absolute  error}                              \\
\hline
$5f^3 7s^2 6d^1 \rightarrow $ &          &            &            &            &            &            &                \\
$5f^3 7s^2 7p^1             $ &    7383  &        387 &       -498 &        -35 &        -33 &          2 &         14     \\
$5f^3 7s^2                  $ &   36159  &        332 &        130 &          4 &          6 &          3 &         16     \\
$5f^3 7s^1 6d^2             $ &   13299  &       -192 &       -154 &         -3 &         -5 &         -1 &        -16     \\
$5f^3 7s^1 6d^1 7p^1        $ &   17289  &        144 &       -621 &        -31 &        -31 &         -1 &         -5     \\
$5f^3 6d^2                  $ &   54892  &       -121 &       -398 &        -14 &        -15 &          1 &        -21     \\
$5f^3 7s^2 6d^1 \rightarrow $ &          &            &            &            &            &            &                \\
$5f^4 7s^2                  $ &   16483  &        176 &        788 &       -723 &          0 &         54 &        187     \\
$5f^4 7s^2      \rightarrow $ &          &            &            &            &            &            &                \\
$5f^4 7s^1 6d^1             $ &   15132  &       -738 &        -87 &         11 &        -11 &        -16 &        -35     \\
$5f^4 7s^1 7p^1             $ &   15016  &         90 &       -443 &        -37 &        -26 &         -1 &         -2     \\
$5f^4 6d^2                  $ &   34022  &      -1287 &       -153 &         28 &        -13 &        -26 &        -62     \\
$5f^4 6d^1 7p^1             $ &   32341  &       -794 &       -457 &        -11 &        -23 &        -17 &        -39     \\
$5f^3 7s^2 6d^1 \rightarrow $ &          &            &            &            &            &            &                \\
$5f^2 7s^2 6d^2             $ &    3774  &       3096 &       -748 &        -17 &        -17 &         90 &        -96     \\
$5f^2 7s^2 6d^2 \rightarrow $ &          &            &            &            &            &            &                \\
$5f^2 7s^2 6d^1 7p^1        $ &   12646  &       -441 &       -626 &        -16 &        -15 &         -5 &          0     \\
$5f^2 7s^2 6d^1             $ &   42638  &       -498 &        155 &         24 &         25 &         -5 &          1     \\
$5f^2 7s^1 6d^3             $ &   10697  &        608 &       -240 &        -10 &        -10 &         13 &          1     \\
$5f^2 7s^1 6d^2 7p^1        $ &   19319  &        390 &       -826 &        -26 &        -26 &          6 &          0     \\
$5f^3 7s^2 6d^1 \rightarrow $ &          &            &            &            &            &            &                \\
$5f^1 7s^2 6d^3             $ &   29597  &      11666 &      -1526 &       -896 &       -104 &        466 &         48     \\
$5f^1 7s^2 6d^3 \rightarrow $ &          &            &            &            &            &            &                \\
$5f^1 7s^2 6d^2 7p^1        $ &   18141  &      -1367 &       -778 &         46 &         49 &         -2 &         -2     \\
$5f^1 7s^2 6d^2             $ &   49158  &      -1355 &        173 &         70 &         73 &         -3 &         -2     \\
$5f^1 7s^1 6d^4             $ &    7584  &       1655 &       -331 &        -39 &        -40 &         22 &         14     \\
$5f^1 7s^1 6d^3 7p^1        $ &   21154  &        779 &      -1055 &        -11 &        -11 &         16 &         10     \\
$5f^3 7s^2 6d^1 \rightarrow $ &          &            &            &            &            &            &                \\
$5f^5                       $ &  100840  &        430 &       1453 &      -1860 &         22 &        105 &        291     \\
\hline
\end{tabular}
\end{table}

\begin{table}
\setcaptionmargin{0mm} \onelinecaptionsfalse
\captionstyle{flushleft}
\caption{\label{CGRECP}
Transition energies from all-electron DC and different GRECP 
calculations 
of the lowest-lying states of the mercury atom and its ions in 
the $[7,9,8,6,7,7]$ correlation basis set from~\cite{Hg} 
for the 12 and 34 correlated 
electrons\protect\footnote{This number is smaller by one or two for Hg$^+$ or 
Hg$^{2+}$ ions, respectively.} 
by the RCC method. All values are in cm$^{-1}$.}
\begin{tabular}{@{}lrrrr}
\hline
 State (leading              & DC      & DC      & GRECP     & CGRECP     \\
 conf., term)                & 34e-RCC & 12e-RCC & 12e-RCC   & 12e-RCC    \\
\hline                                                     
 $5d^{10} 6s^2 (^1S_0) \rightarrow$ &  &         &           &            \\
 $5d^{10} 6s^1 6p^1 (^3P_0)$ &  37471  &  37208  &  37244    &  37742     \\
 $5d^{10} 6s^1 6p^1 (^3P_1)$ &  39318  &  38992  &  39025    &  39573     \\
 $5d^{10} 6s^1 6p^1 (^3P_2)$ &  44209  &  43675  &  43710    &  44453     \\
 $5d^{10} 6s^1 6p^1 (^1P_1)$ &  55419  &  54769  &  54780    &  55466     \\
 $5d^{10} 6s^1 (^2S_{1/2})$  &  84550  &  83885  &  83919    &  84774     \\
 $5d^{10} 6s^1 (^2S_{1/2}) \rightarrow$ & &      &           &            \\
 $5d^{10} 6p^1 (^2P_{1/2})$  &  52025  &  51515  &  51559    &  52059     \\
 $5d^{10} 6p^1 (^2P_{3/2})$  &  61269  &  60476  &  60532    &  61320     \\
 $5d^{10} (^1S_0)$           & 151219  & 150132  & 150202    & 151262     \\
\hline
\end{tabular}
\end{table}

\end{document}